\documentclass[conference]{IEEEtran}
\IEEEoverridecommandlockouts

\usepackage{graphicx}

\usepackage{xspace}
\usepackage{amsmath}
\usepackage{caption}
\usepackage{subcaption}
\usepackage{placeins}
\usepackage{balance}
\usepackage{cite}
\usepackage{bm}
\usepackage{amsmath}
\usepackage{array}
\usepackage{setspace} 
\usepackage[nameinlink]{cleveref}
\usepackage{accents}
\usepackage{mathtools}
\usepackage{algorithm}
\usepackage{algpseudocode}
\usepackage{amsthm} 
\usepackage{multirow}
\usepackage{tabularx}
\usepackage{color}
\usepackage{epstopdf}
\usepackage{endnotes}
\usepackage{framed}
\usepackage{float}
\usepackage{cool} 
\usepackage{enumerate} 

\usepackage{pifont}

\Crefname{figure}{Fig.}{Figs.}

\usepackage{mathrsfs}
\usepackage[justification=centering]{caption}
\allowdisplaybreaks[4]

\newcolumntype{P}[1]{>{\centering\arraybackslash}p{#1}}

\DeclareUnicodeCharacter{2212}{-}

\usepackage{color}

\def\BibTeX{{\rm B\kern-.05em{\sc i\kern-.025em b}\kern-.08em
        T\kern-.1667em\lower.7ex\hbox{E}\kern-.125emX}}

\usepackage{color}
\usepackage[left=1.62cm,right=1.62cm,top=1.9cm]{geometry}
\usepackage{titlesec}

\def\BibTeX{{\rm B\kern-.05em{\sc i\kern-.025em b}\kern-.08em
    T\kern-.1667em\lower.7ex\hbox{E}\kern-.125emX}}
\begin{document}

\title{Tailoring Semantic Communication at Network Edge: A Novel Approach Using Dynamic Knowledge Distillation}

\author{\IEEEauthorblockN{Abdullatif Albaseer, Mohamed Abdallah}
\IEEEauthorblockA{Division of Information and Computing Technology, College of Science and Engineering,
\\Hamad Bin Khalifa University, Doha, Qatar \\
\{aalbaseer, moabdallah\}@hbku.edu.qa}
}
\maketitle

\begin{abstract}
Semantic Communication (SemCom) systems, empowered by deep learning (DL),  represent a paradigm shift in data transmission. These systems prioritize the significance of content over sheer data volume. However, existing SemCom designs face challenges when applied to diverse computational capabilities and network conditions, particularly in time-sensitive applications. A key challenge is the assumption that diverse devices can uniformly benefit from a standard, large DL model in SemCom systems. This assumption becomes increasingly impractical, especially in high-speed, high-reliability applications such as industrial automation or critical healthcare. Therefore, this paper introduces a novel SemCom framework tailored for heterogeneous, resource-constrained edge devices and computation-intensive servers. Our approach employs dynamic knowledge distillation (KD) to customize semantic models for each device, balancing computational and communication constraints while ensuring Quality of Service (QoS). We formulate an optimization problem and develop an adaptive algorithm that iteratively refines semantic knowledge in edge devices, resulting in better models tailored to their resource profiles. This algorithm strategically adjusts the granularity of distilled knowledge, enabling devices to maintain high semantic accuracy for precise inference tasks, even under unstable network conditions. Extensive simulations demonstrate that our approach significantly reduces model complexity for edge devices, leading to better semantic extraction and achieving the desired QoS.
\end{abstract}

\IEEEpeerreviewmaketitle

\section{Introduction}
\label{introduction}
The integration of artificial intelligence (AI) has become a crucial factor in the evolution of modern network systems. Leveraging advanced deep learning (DL) methodologies, complex computational demands are now being met with increased efficiency. However, traditional communication frameworks face challenges in meeting the diverse service demands due to the scarcity of wireless resources in various applications. This necessitates a significant shift in developing next-generation 6G networks. A transformation is envisioned from the conventional bit-centric approach to a more intelligent, AI-driven semantic communication (SemCom) paradigm \cite{shi2021semantic}.  
SemCom epitomizes the transformation towards goal-oriented and task-specific information exchange. It prioritizes the meaning and relevance of data, marking a groundbreaking phase for mission-critical applications. In this new era, the efficiency and reliability of the content conveyed hold equal significance to the precision of the signal transmission, ensuring that the communication is not just accurate but also contextually rich and effective for the intended goals \cite{xie2021task, lan2021semantic,zhang2021toward}. 
The objective is extracting and conveying only the most essential data (semantic contents) tailored to the receiver’s needs, thus aligning communication with intended task execution \cite{9518240}. DL-empowered semantic systems capitalize on this advance, leveraging neural networks (NNs) prowess in extracting and relaying semantic richness from different environments \cite{lan2021semantic,liu2021semantics,yang2021semantic}. These NNs bring unparalleled agility in managing pragmatic communication tasks, surpassing the capabilities of conventional communication techniques \cite{shi2021knowledge,xie2021task}.

Recent investigations in SemCom have primarily concentrated on enhancing data transmission to build reliable systems. These systems are designed to intelligently parse and convey semantics, accounting for an extensive array of data modalities \cite{shi2021new,liu2021semantics,xie2021task}. A notable creation in this domain is the integration of NNs to serve as both semantic encoders and decoders, showcasing enhanced performance \cite{liew2021economics,yang2021semantic}. 
Building on the principles of transfer learning, Domain Adaptation (DA) methodologies strive to reconcile disparities between transmitter and receiver domains—a task of particular relevance to the objectives, given the often-observed discrepancies between training data and real-world scenarios \cite{zhang2022deep}. Cutting-edge DA strategies focus on achieving domain invariance, ensuring fidelity in information retention, and the synthetic generation of target-domain data elements \cite{rozantsev2018beyond}.

Focusing on optimizing the SemCom with respect to wireless resource constraints, Yan et al. \cite{9763856} proposed an approach to enhance reliability and efficiency in low signal-to-noise conditions. They introduced the concept of semantic spectral efficiency and optimized channel allocation and semantic symbol transmission for text-based communication. The authors in \cite{yan2022qoe} explored a quality of experience (QoE) for assessing task-specific SemCom networks (TOSCN), incorporating semantic transmission rates and semantic similarity scores to evaluate service quality and task fulfillment. The goal is to create a semantic-aware allocation system that improves TOSCN's Quality of Service (QoS). Meanwhile, Le et al. \cite{xia2022wireless} studied the problem of resource management in different networks having heterogeneous background knowledge base (BKB), proposing 'system throughput in messages' as a new performance metric. A heuristic algorithm addressed user association and bandwidth distribution in these semantically enabled networks. 
The work in \cite{10122232} introduced a dynamic resource allocation strategy for TOSCN employing deep reinforcement learning (DRL) to prioritize data based on its semantic content for limited resource usage. They explored the interplay between semantic information and task efficacy, presenting a DDPG-driven model for image classification tasks to optimize semantic compression, power, and bandwidth allocation. 

Despite these considerable efforts, i.e., \cite{shi2021new, liu2021semantics,xie2021task,strinati20216g,9763856,yan2022qoe,xia2022wireless,10122232}, there are still significant challenges. A major issue is the neglect of diverse local resource capacities of individual devices and the central coordinating server. This is especially crucial, especially where computational and network constraints vary greatly. Additionally, the requirements for time-sensitive and fault-intolerant systems, which demand reliable and timely task execution and communication, have not been adequately addressed. Most crucially, the heterogeneity and limitations at the edge device level have been significantly overlooked. The general assumption that diverse devices can uniformly benefit from a standard, large NN model in a SemCom system is increasingly impractical, particularly in high-speed, high-reliability applications such as industrial automation or critical healthcare. This highlights an important question for our research: How can we ensure the QoS of SemCom with respect to the heterogeneity of resources and data across devices considering computation, communication, and time constraints?

Motivated by these remarks, this paper aims to develop, optimize, and fine-tune a task-oriented SemCom system. This system caters to the diverse capabilities of heterogeneous, resource-constrained edge devices and computation-intensive servers. Our approach focuses on equipping each device with a customized semantic extraction (SemEx) model that balances computational and communication constraints while meeting strict deadlines and ensuring the required QoS. We leverage the concept of knowledge distillation (KD) and introduce a novel iterative approach to dynamically determine the optimal number of distilled blocks to be transferred from the server to each edge device, ensuring efficient adaptation and performance.
Our contributions are multi-fold and can be delineated as follows:
\begin{itemize}
\item {Formulate an optimization problem considering all aforementioned constraints, followed by an adaptive algorithm that iteratively enhances semantic knowledge within each edge device, considering both computation and communication resources. }
\item Introduce a three-stage dynamic KD-based approach to tailor multi-student semantic models. Our algorithm ensures that each device operates optimally to extract the semantics, aiding the server in efficiently accomplishing the required task.
\item {Conduct extensive simulations; our results show that our proposed approach significantly reduces the model complexity for edge devices without compromising the semantic understanding required for accurate inference tasks.  Additionally, it also reduces the resource consumption required for communication. }
\end{itemize}
The rest of this paper is structured as follows. Section \ref{sec:system} illustrates the system model. In Section \ref{sec:problem_formulation}, we present the problem formulation while introducing our proposed approach in Section \ref{sec:proposed_solution}. The proposed approach is evaluated in Section \ref{sec:performance_evaluation}, and we conclude our work in Section \ref{sec:conclusion}.

\section{System Model}
\label{sec:system}
\begin{figure}
    \centering
    \includegraphics[width=0.7\linewidth]{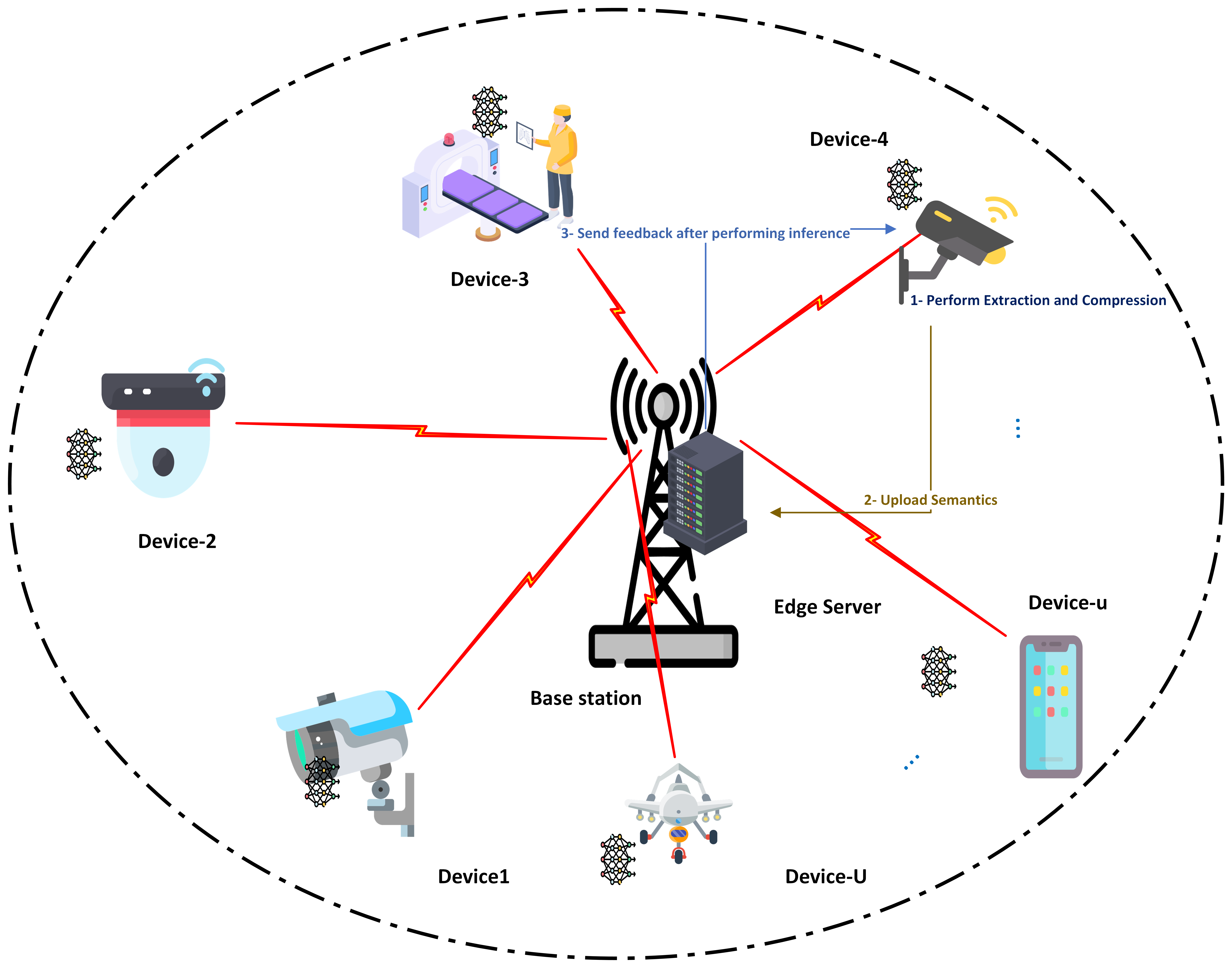}
    \caption{The Task-Oriented SemCom System.}
    \label{fig:sys_model}
\end{figure}
In this paper, as illustrated in \Cref{fig:sys_model}, we consider a distributed SemCom system comprising a set of edge devices, $\mathcal{U}$, and a coordinating server. The server and the devices operate over $\tau$ discrete time slots, each with a defined deadline for task completion. Each device, $u \in \mathcal{U}$, is allocated specific bandwidth and power in each time slot to adapt to varying transmission rates and channel conditions. The system is divided into $J$ task categories, with $U_j$ devices assigned to each task $j$, where $ \sum_{j=1}^{J} U_j = U $. These devices are equipped with vision sensors to capture images and perform SemEx with processing speeds relative to their capabilities. The processed data is then compressed, encoded, and transmitted to the edge server to perform the required task with a required QoS. Subsequently, the edge server provides feedback to the transmitter to update its BKB. 

At the transmitter sides, there is an N-layer deep residual network (ResNet-N), to perform SemEx and compression. In contrast, the edge server utilizes a fully connected (FC) layer as a semantic decoder to perform the inference task. Mathematically speaking, the SemEx process for an input image $ \mathbf{I} $ is defined as:
\begin{equation}
\mathbf{P} = S_{EX}(\mathbf{I}, \theta_{u}),
\end{equation}
where $ S_{EX} $ denotes the SemEx network with its trainable parameters $ \theta_{u}$.
The goal is to select the optimal feature map proxies for semantic information based on their task relevance, established through global average pooling and gradient backpropagation. The weight of each feature map, $ w_{k}^{c}$, reflects its contribution to the class output $c$ as:
$
w_{k}^{c} = \frac{1}{H \times W} \sum_{x=1}^{H} \sum_{y=1}^{W} F_{k}(x, y),
$
where $ F_{k}(x, y) $ is the activation of the $k$-th feature map at location $(x,y)$, with $H$ and $W$ representing the feature map's dimensions.
An Importance List of Feature Maps (ILFM), $ \mathbf{w}^{c}$, is constructed by ranking feature maps according to their absolute weights: $\mathbf{w}^{c} = \text{sort}\left(\left| w_{1}^{c} \right|, ..., \left| w_{N}^{c} \right|\right)$, where $N$ denotes the number of classes in a given task. This approach is crucial for optimizing resource use in time-sensitive and fault-intolerant systems, as in our paper, by prioritizing critical semantics transmission. The semantic compression is outlined as follows:
\begin{equation}
S_{CM}(F^{k},\eta) = 
\begin{cases} 
F^{k}, & \text{if } \left|w_{k}^{c}\right| \geq {\eta} \\
\mathbf{0}, & \text{if } \left|w_{k}^{c}\right| < {\eta},
\end{cases}
\end{equation}
where $\eta \in [0,1)$ is the threshold for compression. It is worth mentioning that balancing compression ratio and task performance is critical, which is managed by the BKB that guides real-time resource allocation to ensure achieving efficient tasks within a given time constraint, as seen later on. The resulting compressed semantics are defined as:
\begin{equation}
\mathbf{M} = S_{CM}(\mathbf{P}, \eta),
\end{equation} where $S_{CM}$ is the semantic compression operation. 
Last, the channel encoding process $ C_{EN} $ is expressed as follows:
\begin{equation}
\mathbf{X} = C_{EN}(\mathbf{M}, \theta_{EN}),
\end{equation}
where $\theta_{EN}$ represents the neural network's trainable parameters.
Due to space constraints and for the sake of brevity, detailed descriptions of receiver-side operations are omitted. These operations involve executing inverse processes, namely $C_{EN}^{-1}$, $S_{CM}^{-1}$, and $S_{EX}^{-1}$. 

For the communication and computation models, the resulting data, post SemEx and compression, has a size $D_u = |\mathbf{I}|$, which remains consistent unless compression is applied. For user $u$ during the $t$-th time slot:
\begin{equation}
\hat{D}_{t}^{u} = (1 - \eta{t}^{u})D_u,
\end{equation}
where $\eta_{t}^{u}$ represents the user's compression ratio for user, $u$, at time slot $t$. Following that, the user's transmission rate is given by:
\begin{equation}
R_{t}^{u} = B_{t}^{u} \log \left(1 + \frac{P_{t}^{u} h_{t}^{u}}{\sigma_{t}^{u^2}}\right),
\end{equation}
with $ B_{t}^{u} $ and $ P_{t}^{u} $ indicating the bandwidth and transmit power for user $u$. The terms $ {\sigma_{t}^{u^2}} $ and $ h_{t}^{u} $ relate to the noise power and channel gain, respectively. Here, the channel gains account for Rayleigh fading, with:
$
h_{t}^{u} = \alpha^{u} g_{t}^{u}, 
$
and large-scale fading expressed as:
$
\alpha^{u} = G^{u} \beta^{u} (d^{u})^{-\varphi^{u}}, 
$ where $G^{u}$ is the constant path loss, $\beta^{u}$ is the shadowing component, $d^{u}$ is the distance between the edge server and the device, and $\varphi^{u}$ is the path loss component. 
Therefor, the time required for user $u$ to upload an extracted semantics, $\hat{D}_{t}^{u}$, is:
\begin{equation}
T_{comm}^{u} = \frac{\hat{D}_{t}^{u}}{R_{t}^{u}},
\end{equation}
and the associated energy consumption:
\begin{equation}
E_{comm}^{u} = T_{comm}^{u} \times P_{t}^{u}.
\end{equation}

Considering the resource heterogeneity among devices, the computation model is defined as follows.
The time required to capture an image, $T_{cap}^{u}$, is defined as:$
T_{cap}^{u} = \frac{L_{pixels}^{u}}{R_{read}^{u} \cdot E_{eff}^{u}},
$
where $L_{pixels}^{u}$ is the number of pixels captured by device, 
$R_{read}^{u}$ is the readout rate of the sensor in device $u$ (pixels per second), and $E_{eff}^{u}$ is the efficiency of the image processing pipeline in device $u$ (a value between 0 and 1). The SemEx time, $T_{ext}^{u}$, is given by $
T_{ext}^{u} = \frac{A(\theta_{u}) \cdot D_u }{f_{comp}^{u}},
$
where $A(\theta_{u})$, and $f_{cmp}^{u}$ represent the model complexity and computational speed, respectively. It is important to highlight that $T_{cap}^{u}$ depends on the hardware itself while $T_{ext}^{u}$ depends mainly on the complexity of the trained model, which we aim to optimize as seen later in Sections III and IV.  The total time, $T_{total}^{u}$, to perform both tasks is defined as:
\begin{equation}
T_{cmp}^{u} = T_{cap}^{u} + T_{ext}^{u},
\end{equation}
and the corresponding energy consumption:
\begin{equation}
E_{total,cmp}^{u} = P_{cap}^{u} \times T_{cap}^{u} + P_{ext}^{u} \times T_{ext}^{u},
\end{equation}
where $P_{cap}^{u}$ and $P_{ext}^{u}$ are the power consumption of device $u$ during image capture and SemEx. It is important to highlight that we modeled the encoding time and energy as a part of the transmission. 

\section{Problem Formulation}
\label{sec:problem_formulation}
Given the system model in Section \ref{sec:system}, we aim to develop, optimize, and fine-tune a task-oriented SemCom system tailored to adapt the diverse capabilities of heterogeneous, resource-constrained edge devices and computation-intensive servers. Specifically, we seek to equip each device with an optimally performing SemEx model that balances computational and communication limitations, adheres to strict deadlines, and ensures the required QoS. 
To achieve this, we take advantage of using the concept of KD \cite{motamedi2019distill}. Specifically, we aim to optimize the accuracies $\Omega = \{\Omega_1, \dots, \Omega_U, \Omega_T\}$ of local student models $\Theta_S = \{\theta_1, \dots, \theta_U\}$ and the server teacher model $\Theta_T$, while attaining the computation and communication constraints imposed by the resource heterogeneity as well as the deadline imposed by a time-sensitive application. The student models (devices models) are trained through the distillation technique that exploits the knowledge from the teacher model $\Theta_T$. This guides us to  formulate the following intricate optimization formulation:
\begin{align}
\textbf{P1:}    \max_{\Theta_S, \Theta_T} \quad & \sum_{u=1}^{U} w_u \Omega_{u}(\theta_{u}) + \lambda \Omega_T(\theta_T)  \\
    \text{s.t.} \quad  C_1: \quad & \Omega_u, \Omega_T  \geq \Omega_{\text{min}}, \quad \forall u \quad \& \quad T, \\
    C_2: \quad& T_{\text{comm}}^{u} + T_{\text{cmp}}^{u} (\theta_{u}) \leq T_{\text{max},u} \quad \forall u, \\
    C_3: \quad& \sum_{u=1}^{U} B_{t}^{u} \leq B, \\
    C_4: \quad& E_{\text{comm}}^{u} + E_{\text{total,cmp}}^{u} (\theta_{u}) \leq E_{\text{budget},u}, \\
    C_5: \quad& D_u(\theta_u) = D_T(\theta_T), \quad \forall u \\
    C_6: \quad& \theta_{u}, \theta_T \in \Theta \quad \forall u. 
\end{align}
In \textbf{P1}, the constraint $C_1$ ensures that both the student and teacher models adhere to a desired level of accuracy, reflecting the minimum QoS requirements. The constraint, $C_2$, ensures adherence to the required sensitive time, $T_{max,u}$. The bandwidth constraint, $C_3$, ensures that the bandwidth utilization by all edge devices does not exceed the total bandwidth. The energy constraint, $C_4$, is imposed to keep the energy consumption for communication and computation within the allowed budget. The condition $D_u(\theta_u) = D_T(\theta_T)$ in $C_5$ ensures that both models are trained on congruent data distribution. The set $\Theta$ in $C_6$ defines the feasible space of all model parameters. Finally, the weighting parameters $w_u$ and $\lambda$ balance the relative importance of individual model accuracies and the overall system performance. Precisely, $w_u$ adjusts the contribution of each student model's SemEx accuracy to the overall objective, enabling customization of the solution to prioritize specific devices or tasks as needed. 
We note that \textbf{P1} is inherently complex and is proven to be NP-hard, given the non-convexity of the accuracy function with respect to the model parameters.

\section{Proposed Solution}
\label{sec:proposed_solution}
Addressing the dynamic and stochastic nature of the problem, \textbf{P1}, particularly the non-stationary conditions, variability in data distributions across devices, and fluctuations in resource availability, necessitates an adaptive and more robust solution. The static and traditional distillation process (i.e., distilled $N_{\text{Blocks}}$) from the server model $\Theta_T$ does not fit well, mainly due to high variations in local resources. It fails to capture the strict requirements of real-time system adaptability and does not consider the possibility of sudden shifts in resource allocations or data characteristics. \textbf{In response,} we propose an iterative pre-deployment approach that dynamically fine-tunes the number of distilled blocks before deployment while attaining all aforementioned constraints. This leads to reformulating \textbf{P1} by introducing a new decision variable, $N_{\text{Bdistilled}}$ along with the decision variables $\Theta_S, \Theta_T$, to adopt all these challenges, ensuring that each model at the edge is fine-tuned to operate within the specified resource and operational constraints. The problem is reformulated as:

\begin{align}
   \textbf{P2:}\quad \max_{\Theta_S, \Theta_T, N_{\text{Bdistilled}}} \quad & \sum_{u=1}^{U} w_u \Omega_{u}(\theta_{u}, N_{\text{Bdistilled}, u}) + \lambda \Omega_T(\theta_T)  \\
    \text{s.t.} \quad & C_1 \quad \textit{to} \quad C_6 \quad \textit{in} \quad \textbf{P1} \\
    & C7: \quad N_{\text{Bdistilled}, u} \leq N_{\text{Blocks}},
\end{align}
Solving \textbf{P2} still poses a considerable challenge due to its intrinsic intractability, primarily how to directly determine the optimal number of distilled blocks from the server to the device model. Additionally, the nature of an iterative loss function, which is crucial for refining and optimizing models, further requires an iterative solution. Therefore, we propose an efficient pre-deployment iterative solution to address these complexities by recursively adjusting the number of distilled blocks considering all aforementioned constraints.
Our solution introduces an iterative algorithm to distill the knowledge dynamically while attaining all constraints. Mathematically: for a server function $F_T: \mathcal{X} \to \mathcal{Y}$ with parameters $\theta_T$, and a corresponding edge device function $F_u: \mathcal{X} \to \mathcal{Y}$ with parameters $\theta_u$, our objective is to iteratively minimize the divergence $D(F_T(x)||F_u(x))$ for each device that fits the system requirements and the local capabilities. This divergence typically instantiated as the Kullback-Leibler (KL) Divergence, quantifies the disparity in semantic content representation between server and edge devices' models, including three stages in our proposed approach detailed as follows:
\subsubsection{{Stage 1: Establishing a Baseline with Simplified Task or Model}}
\label{subsec:initial_stage}
 \noindent \textbf{{Baseline Establishment}}: {Initiated with $\theta_{T_{\text{init}}}$, this stage sets a foundational groundwork, easing the device's model into the complexity of the SemEx task at the server.}\newline \textbf{{Loss Function Specification}}: {The KL divergence loss function, softened by a temperature parameter $\zeta$, quantifies the initial learning gap:}
    \begin{equation}
\mathcal{L}_{\text{KL}}(\theta_u, \theta_{T_{\text{init}}}) = \sum_i \frac{e^{F_u(x; \theta_u)_i / \zeta}}{\sum_j e^{F_u(x; \theta_u)_j / \zeta}} \log \left( \frac{\frac{e^{F_u(x; \theta_u)_i / \zeta}}{\sum_j e^{F_u(x; \theta_u)_j / \zeta}}}{\frac{e^{F_{T_{\text{init}}}(x; \theta_{T_{\text{init}}})_i / \zeta}}{\sum_j e^{F_{T_{\text{init}}}(x; \theta_{T_{\text{init}}})_j / \zeta}}} \right) \nonumber
    \end{equation}\newline \textbf{{Optimization Initiation}}: {This phase involves seeking an optimal parameter set $\theta_u^*$ that reduces the initial model gap:}
    $
    \theta_u^* = \arg \min_{\theta} \mathcal{L}_{\text{KL}}(\theta_u, \theta_{T_{\text{init}}}).
    $
\subsubsection{{Stage 2: Progressive Transition to Full Task Complexity}}
\label{subsec:transition_stage}
\noindent \textbf{{Transitional Loss Function}}: {by integrating a dynamic shift parameter $\alpha$, this loss function bridges the initial and final edge model stages:}
   $
\mathcal{L}_{\text{trans}}(\theta_u, \theta_T, \theta_{T_{\text{init}}}) = \alpha \mathcal{L}_{\text{KL}}(\theta_u, \theta_{T_{\text{init}}}) + (1 - \alpha) \mathcal{L}_{\text{KL}}(\theta_u, \theta_T)
   $\newline\textbf{{Dynamic Optimization Strategy}}: {The aim is to iteratively adjust $\theta_u^*$ to mirror the increasing SemEx task complexity:}
    $
    \theta_u^* = \arg \min_{\theta} \mathcal{L}_{\text{trans}}(\theta_u, \theta_T, \theta_{T_{\text{init}}})
    $
\subsubsection{{Stage 3: Mastery of Original Task Complexity}}
\label{subsec:final_stage}
\noindent \textbf{{Final Loss Function}}: {Utilizes the original KL divergence, reflecting the knowledge and complexity of the final server model $\theta_T$:}
    \begin{equation}
\mathcal{L}_{\text{KL}}(\theta_u, \theta_T) = \sum_i \frac{e^{F_u(x; \theta_u)_i / \zeta}}{\sum_j e^{F_u(x; \theta_u)_j / \zeta}} \log \left( \frac{\frac{e^{F_u(x; \theta_u)_i / \zeta}}{\sum_j e^{F_u(x; \theta_u)_j / \zeta}}}{\frac{e^{F_T(x; \theta_T)_i / \zeta}}{\sum_j e^{F_T(x; \theta_T)_j / \zeta}}} \right) \nonumber
    \end{equation}\newline \textbf{{Ultimate Optimization Objective}}: {Seeks the optimal $\theta^*$ that achieves the highest fidelity in mimicking the teacher model:}
    $
    \theta^* = \arg \min_{\theta} \mathcal{L}_{\text{KL}}(\theta_u, \theta_T).
    $
{The goal at every stage is to gradually improve the model of each device so it closely approaches the performance of the larger model on the server side:}
{\begin{equation}
\footnotesize
\theta_u^* = \arg \min_{\theta} \left( \mathcal{L}_{\text{KL}}(\theta_u, \theta_{T_{\text{init}}}) + \mathcal{L}_{\text{trans}}(\theta_u, \theta_T, \theta_{T_{\text{init}}}) + \mathcal{L}_{\text{KL}}(\theta_u, \theta_{T}) \right) \nonumber
\end{equation}}

{As detailed in Alg. \ref{alg:framework}, our approach adaptively refines the distilled blocks in response to real-time feedback on system performance and resource limitations. This feedback loop is integral to the optimization process, ensuring that each iteration balances between model accuracy, $\Omega_{u}$, and the constraints imposed by computational capabilities and energy budgets of each device.}

\begin{algorithm}[t]
\footnotesize
\caption{{Iterative Distillation for Heterogeneous Device Optimization}}
\begin{algorithmic}[1]
\State \textbf{Inputs:} {$K$, $\Omega_{\text{min}}$, $T_{\text{max}}$, $E_{\text{budget}}$, $N_{\text{blocks}}$, $\epsilon$}

\State \textbf{Initialization:} 
\State {$N_{\text{distilled}} \gets \{\text{for each } u:  N_{\text{blocks},u} \}$}

\For{{$k \gets 1$ to $K$}}
    \State {$prev\_N_{\text{distilled}} \gets N_{\text{distilled}}$}
    
    \ForAll{{$u \in \text{users}$}}
    \State Apply our proposed three-stage process to get $\theta_u$
        \State {Assess $\Omega_u$}
        \State Compute $T_u = T_{comm}^u + T_{cmp}^u$, \State Compute $E_u = E_{comm}^u + E^u_{cmp}$ 
        \If{{$T_u \leq T_{\text{max},u}$ \textbf{and} $E_u \leq E_{\text{budget},u}$}}
            \If{{$\Omega_u < \Omega_{\text{min}}$ \textbf{and} $N_{\text{distilled},u} < N_{\text{blocks},u}$}}
                \State {Increase distilled blocks by 1 for user $u$}
                \State {Break;}
            \ElsIf{{$\Omega_u \geq \Omega_{\text{min}}$ \textbf{and} $N_{\text{distilled},u} > 1$}}
                \State {Decrease distilled blocks by 1 for user $u$}
                \State {Break;}
            \EndIf
        \ElsIf{{$T_u > T_{\text{max},u}$ \textbf{or} $E_u > E_{\text{budget},u}$}}
            \State {Decrease distilled blocks by 1 if possible}
            \State {Break;}
        \EndIf
    \EndFor
    
\EndFor

\State \textbf{return} {$\Theta_T,$ $\Theta_S$ with optimized  $N_{\text{distilled}}$}
\end{algorithmic}
\label{alg:framework}
\end{algorithm}

\section{Performance Evaluation}
\label{sec:performance_evaluation}
\textbf{Experimental Setup: } we evaluate our proposed approach using the CIFAR-100 dataset. We implemented a ResNet-18 architecture as the teacher model to distill knowledge into devices' student models with an optimized number of blocks. We set the number of users as $U = 10$, the minimum quality threshold as $\Omega_{\min} = 0.8$, and the power range for each user $P_t^u$ within $[0.2, 0.5]$ watts. The total available bandwidth is 10 MHz, and the CPU frequency ranges from 0.5 to 2 GHz.
Edge devices are simulated with varying computational capacities, reflecting the heterogeneous nature of real-world deployments. The optimization process, constrained by the devices' energy budgets and the maximum allowable inference time, aimed to maximize accuracy while ensuring resource efficiency. The learning rate for model training was fixed at $0.001$, and the optimization algorithm was executed until convergence was observed.

\begin{figure}[t]
    \centering
\includegraphics[width=0.8\linewidth]{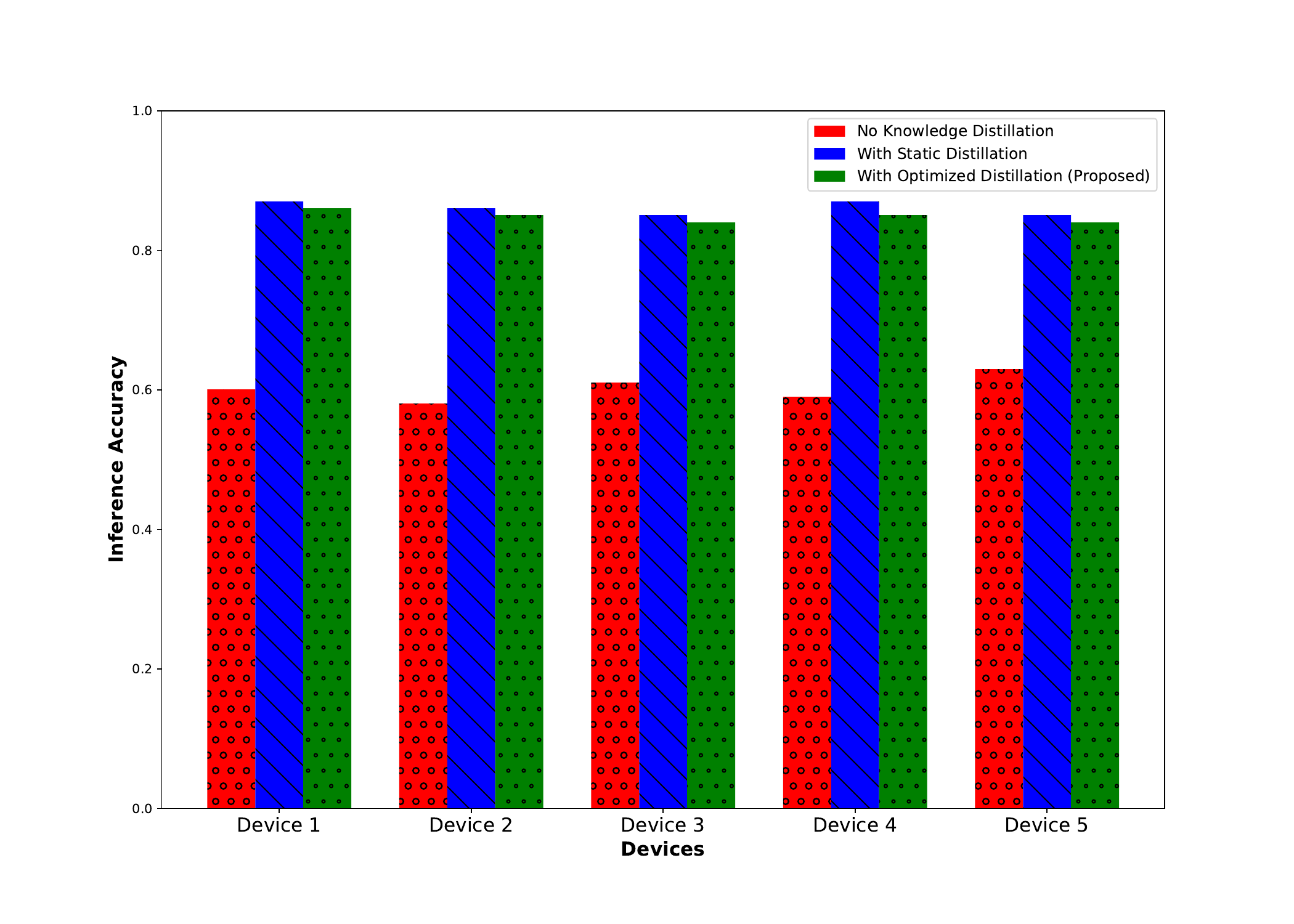}
    \caption{Inference accuracy of selected five devices for both the proposed approach and the baselines.}
    \label{fig:acc_baseline_compare}
\end{figure}

\begin{figure}[t]
    \centering
\includegraphics[width=0.8\linewidth]{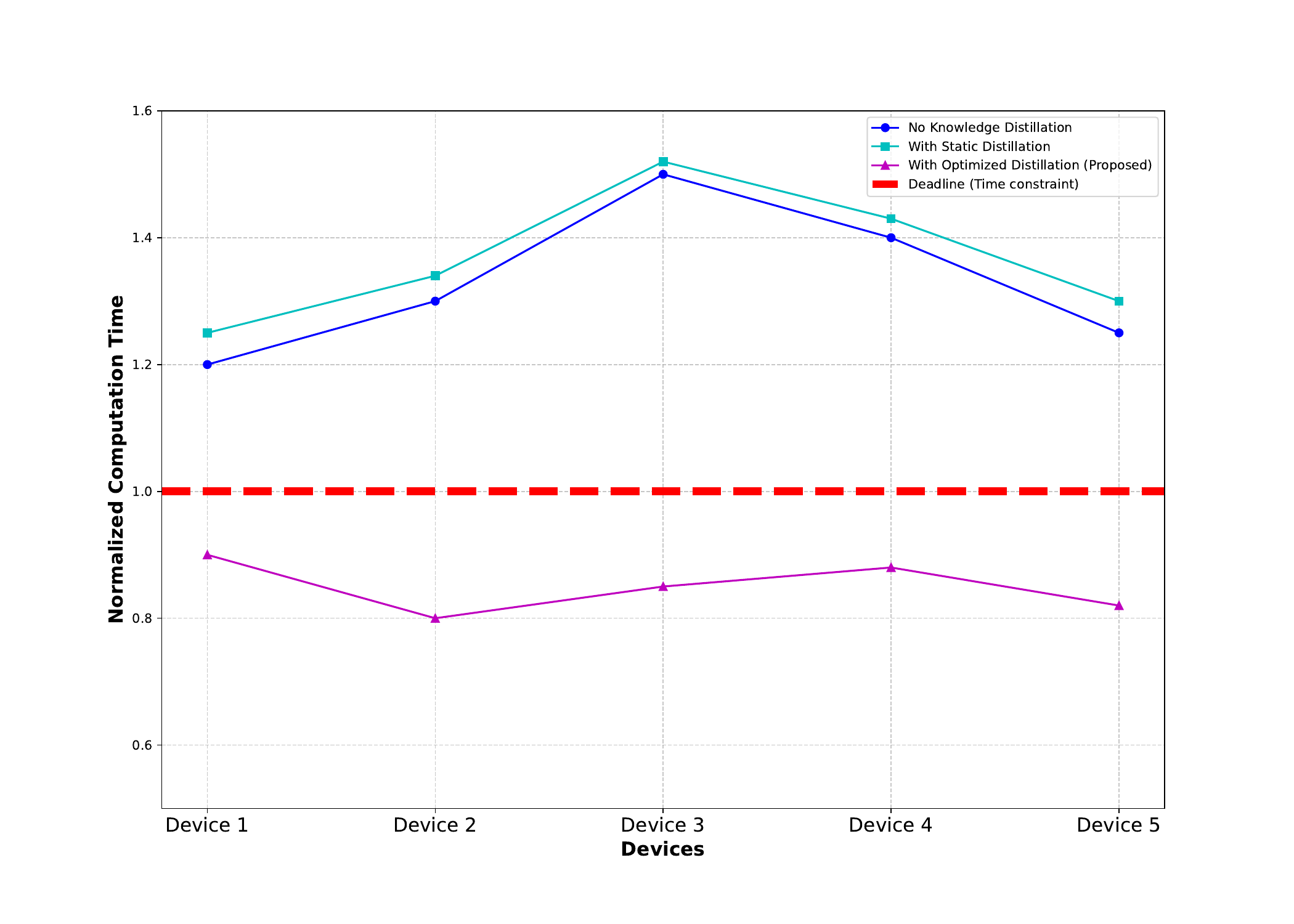}
    \caption{Computational Complexity for both the proposed approach and the baselines.}
\label{fig:normalized_computation_time_compare}
\end{figure}

\textbf{Results and Discussion: }
\Cref{fig:acc_baseline_compare} presents the inference accuracy of the semantic from first five devices across three scenarios: SemCom without KD (where all devices have the same model architecture and are trained using only their local data \cite{xia2022wireless, 10122232}), SemCom with static distillation (where all devices inherit the same number of blocks from the server model), and our proposed approach, in which the number of distilled blocks is optimized based on time, computation, and communication constraints, as well as the required QoS. The results clearly show that our optimized approach greatly improves device performance, performing better than the non-distillation method and almost as well as the static-distillation approach, but with less computational complexity, as shown in \ref{fig:normalized_computation_time_compare}. This results from our proposed iterative approach, which recursively trains the SemEx model using an optimized number of blocks to ensure the required QoS.

In \Cref{fig:normalized_computation_time_compare}, we present the normalized computational time for the same five devices. From this figure, we note that the conventional SemCom scenario without distillation results in the highest computational complexity due to the complexity of the model used. In contrast, the SemCom scenario with static distillation exhibits a similar trend but with a slight reduction in time, thanks to the more efficient model resulting from the distillation. However, our proposed optimized distillation method demonstrates outstanding performance, revealing the lowest computational time. This is due to the iterative optimization process, which seeks to identify the most appropriate blocks during the training phase.

\begin{figure}[t]
    \centering
\includegraphics[width=0.8\linewidth]{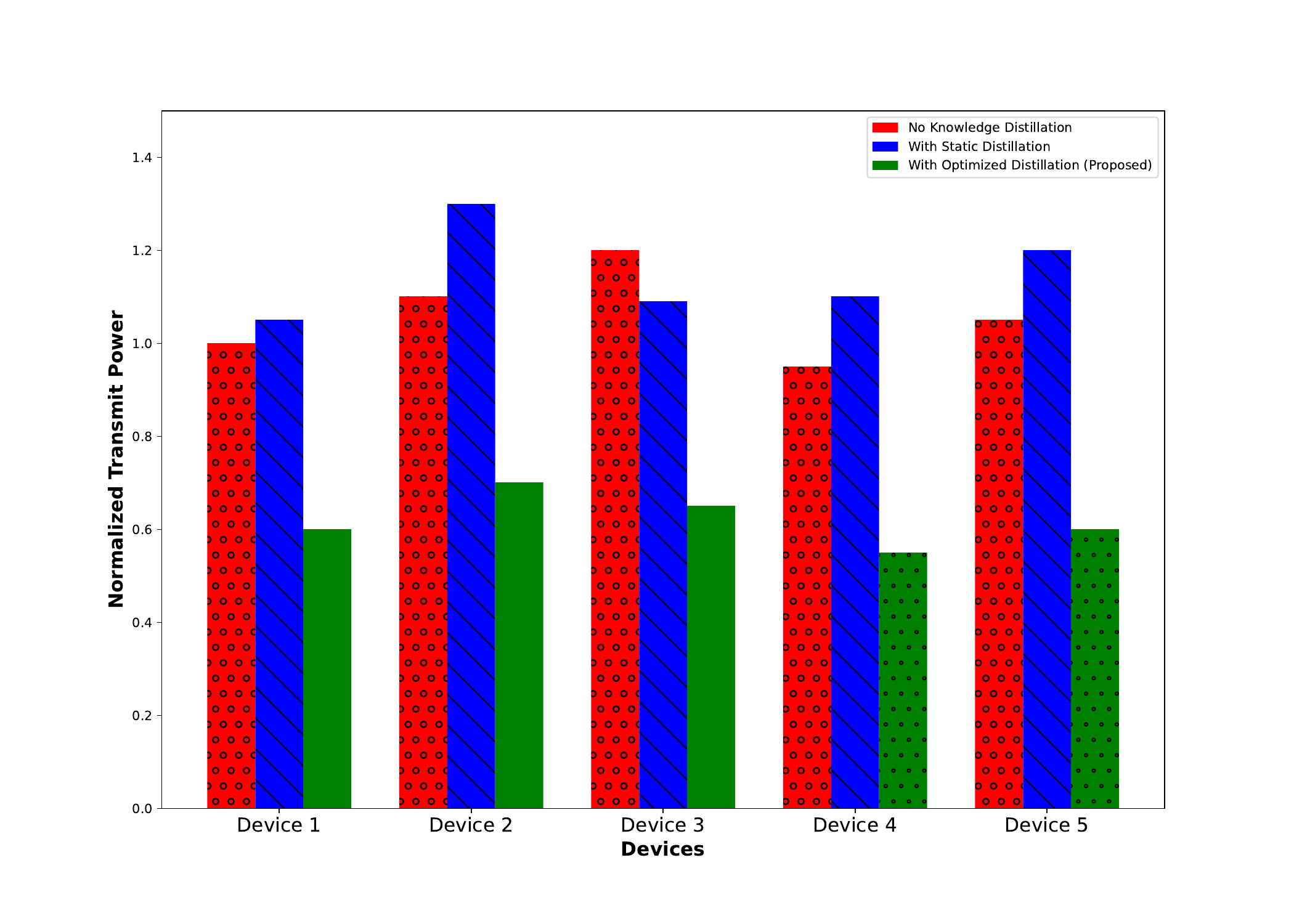}
    \caption{Normalized Transmit Power Consumption of selected five devices for both the proposed approach and the baselines.}
    \label{fig:normalized_transmit_power}
\end{figure}

\Cref{fig:normalized_transmit_power} displays the normalized transmit power consumption of the same five devices across three scenarios. The results indicate that our proposed method consumes the least power. This stems from the fact that each local model learns well to transmit only the most critical semantics, achieved in less time and with reduced complexity. Consequently, this allows more time for the uploading task, resulting in decreased power requirements.

\section{Conclusion}
\label{sec:conclusion}
This paper presented a novel framework for optimizing SemCom systems through dynamic KD, tailored explicitly for heterogeneous edge devices operating under varying computational and network constraints. Our proposed approach is based on a multi-student dynamic distillation technique that addresses the challenge of resource heterogeneity and ensures that the tailored distilled SemEx models maintain the required QoS and computation and communication resources, which is crucial for time-sensitive and fault-intolerant systems.
Through iterative optimization, we demonstrated that the adaptive distillation process could significantly reduce the model complexity for edge devices without compromising the semantic understanding required for accurate inference tasks. The simulation results showed that our approach significantly improved semantic accuracy and reduced network communication overhead.

\bibliographystyle{IEEEtran}
\bibliography{ref}

\end{document}